\documentclass[utf8]{article}
\usepackage{url,hyperref,lineno,microtype,subcaption}
\usepackage[onehalfspacing]{setspace}
\usepackage{graphicx}

\usepackage{natbib}
\begin{document}

\title{Automated Detection of Dolphin Whistles with Convolutional Networks and Transfer Learning} 

\author{Burla Nur Korkmaz, Roee Diamant, Gil Danino, Alberto Testolin}
\date{November 2022}
\maketitle

\begin{abstract}

Effective conservation of maritime environments and wildlife management of endangered species require the implementation of efficient, accurate and scalable solutions for environmental monitoring.
Ecoacoustics offers the advantages of non-invasive, long-duration sampling of environmental sounds and has the potential to become the reference tool for biodiversity surveying. However, the analysis and interpretation of acoustic data is a time-consuming process that often requires a great amount of human supervision.
This issue might be tackled by exploiting modern techniques for automatic audio signal analysis, which have recently achieved impressive performance thanks to the advances in deep learning research.
In this paper we show that convolutional neural networks can indeed significantly outperform traditional automatic methods in a challenging detection task: identification of dolphin whistles from underwater audio recordings.
The proposed system can detect signals even in the presence of ambient noise, at the same time consistently reducing the likelihood of producing false positives and false negatives. Our results further support the adoption of artificial intelligence technology to improve the automatic monitoring of marine ecosystems.
\end{abstract}

\section{Introduction}

Systematic monitoring of marine ecosystems is a key objective to promote sustainability and guarantee natural preservation. Developing and testing innovative monitoring systems is thus rapidly becoming a priority in research agendas, and modern technologies have already shown great potential to advance our understanding of marine communities and their habitat \citep{danovaro2016implementing}.

Acoustic approaches are widely used to investigate underwater activity thanks to their ability to detect and classify sensitive targets even in low visibility conditions; moreover, passive acoustic technologies (e.g., hydrophones) allow to perform non-invasive continuous monitoring without interfering with biological processes \citep{sousa2013review}. Notably, most species of marine mammals are acoustic specialists that rely on sounds for communication, reproduction, foraging and navigational purposes. Here we focus on the task of detecting whistles generated by bottlenose dolphins (\textit{Tursiops truncatus}), which can produce a remarkable variety of sound calls for communication purposes (for a review, see \cite{janik2013communication}).

Traditional bioacoustic tools to detect odontocete vocalizations typically rely on template matching or algorithmic analysis of audio spectrograms. For example, in the reference approach pursued by \cite{gillespie2013automatic} three noise removal algorithms are first applied to the spectrogram of sound data, and then a connected region search is conducted to link together sections of the spectrogram which are above a pre-determined threshold and close in time and frequency. A similar technique exploits a probabilistic Hough transform algorithm to detect ridges similar to thick line segments, which are then adjusted to the geometry of the potential whistles in the image via an active contour algorithm \citep{serra2020active}. Other algorithmic methods aim at quantifying the variation in complexity (randomness) occurring in the acoustic time series containing the vocalization, for example by measuring signal entropy \citep{siddagangaiah2020automatic}.

Nevertheless, automatic environmental monitoring can nowadays be made much more efficient thanks to the deployment of surveying techniques based on Artificial Intelligence. Indeed, recent work has shown that machine learning has the potential to identify signals in large data sets with greater consistency than human analysts, leading to significant advantages in terms of accuracy, efficiency and cost \citep{ditria2022artificial}.
In particular, deep learning approaches based on Convolutional Neural Networks (CNN) \citep{lecun1995convolutional} have been applied to detection of whales vocalizations, producing false-positive rates that are orders of magnitude lower than traditional algorithms, while substantially increasing the ability to detect calls \citep{jiang2019whistle,shiu2020deep}. Deep learning has also been used to automatically classify dolphin whistles into specific categories \citep{li2021automated} and to extract whistle contours by exploiting peak tracking algorithms \citep{li2020learning} or by training CNN-based semantic segmentation models \citep{jin2022semantic}.

Here we further demonstrate the advantage of deep learning models over alternative algorithmic approaches by testing the detection capability of convolutional neural networks on a large-scale dataset of recordings, collected in a series of sea experiments and carefully tagged by human experts. We show that the performance of deep learning models dramatically exceeds that of traditional algorithms, and we further show that transfer learning \citep{pan2009survey} from pre-trained models is a promising way to further improve detection accuracy. The complete dataset of dolphin recordings collected for this study is stored on a cloud server and made publicly available to download \citep{DataRepo}.

\section{Method}

\subsection{Dataset}
\begin{figure}[]
    \centering
    \includegraphics[width=0.7\textwidth]{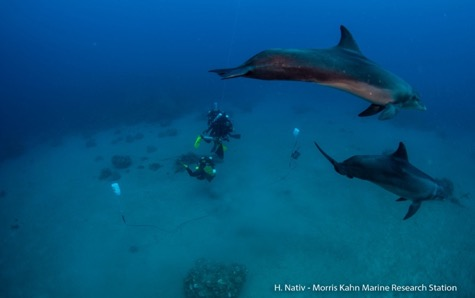}
    \caption{The deployed acoustic recorder with dolphins inspecting the operation. Picture taken from the Eilat deployment site at depth of 50~m.}
    \label{fig:recorder}
\end{figure}

We created a large-scale database of sound recordings by exploiting a self-made acoustic recorder that comprised a Raspberry Pi-Nano, a sound card sampling at 96~kHz@3B, a pre-amplifier, a battery set, two Geospectrum M18 hydrophones, and a custom made housing. The recorder was anchored by scuba divers at depth of 50~m roughly 200~m from the dolphin's reef in Eilat, Israel. Using floats, the hydrophones were set to hang 1.5~m above the seabed. A picture from the deployment is shown in Fig.~\ref{fig:recorder}. The recorder was made to continuously log \textit{flac} files for 27 days during the Summer period of year 2021: once recovered, the data passed a quality assurance (QA) procedure to remove sporadic cut-offs and extensive noise periods. The QA involved canceling of noise transients by wavelet denoising, and identifying and discarding cut-off events by thresholding and bias removal.

\subsection{Data pre-processing and data tagging}
\begin{figure}[]
    \centering
    \includegraphics[width=1\textwidth]{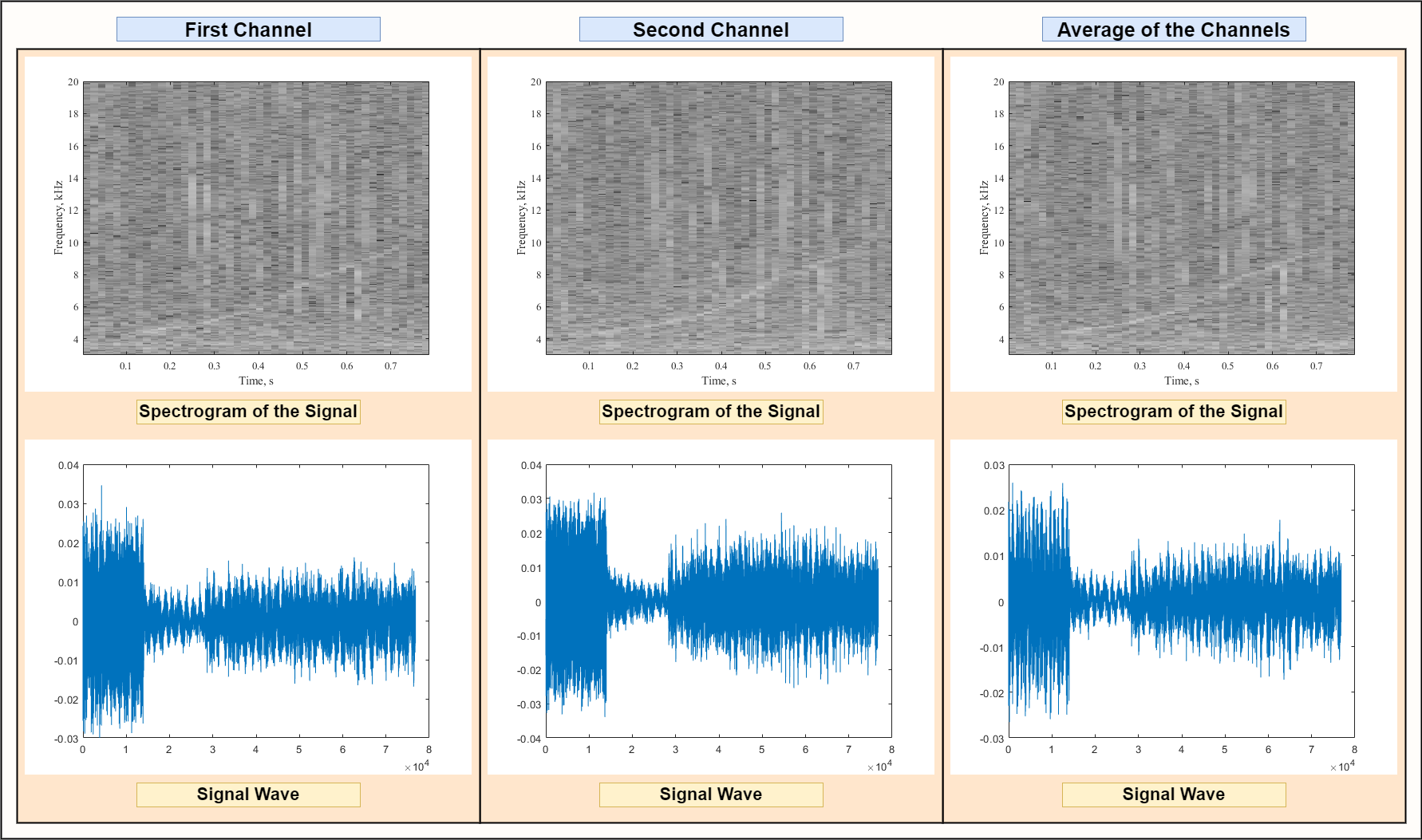}
    \caption{Visualization of the spectrogram (top panels) and raw audio data (bottom panels) of a representative sample containing a dolphin whistle (curved line in the time-frequency plots). Our detection system receives as input the average of the two recording channels.}
    \captionsetup{justification=centering}
    \label{fig:signal_channelling}
\end{figure}

The data passed through a bandpass filter of range 5~kHz-20~kHz to fit most dolphins' whistle vocalizations, and through a whitening filter designed to correct for ripples in the hydrophone's open circuit voltage response and the sound card's sensitivity. Recorded audio files consisted of 2 channels, which were averaged before creating the spectrograms in order to reduce noise (see example in Fig.~\ref{fig:signal_channelling}).
Our pre-processing pipeline also removed signal outliers based on their length, using the quartiles-based Tukey method \cite{tukey1949comparing}. This resulted in discarding signals longer than 0.78 seconds and shorter than 0.14 seconds.

Spectrograms of dolphin whistles were then created by calculating the short-time fast Fourier transform of the signal using MATLAB’s \textit{spectrogram} function from the digital signal processing toolbox, using a Blackman function window with 2048 points, periodic sampling and a hop size obtained by multiplying the window length by 0.8. Subsequent spectrograms were calculated by shifting the signal window by 0.4 seconds. Spectrogram images were finally produced by applying a gray-scale colormap, converting the frequency to kHz and the power spectrum density to dB and limiting the y-axis between 3 and 20 kHz to focus on the most relevant frequency range.

Spectrograms were then manually tagged by one human expert in two phases: 1) marking tagging and 2) validation tagging. The former involved accurate annotation of 5~seconds spectrograms over 10 days of data collection, in order to train a preliminary version of a deep learning classifier that was then used to select new portions of recordings containing putative dolphin sounds. This allowed to more efficiently tag the remaining data during the validation tagging phase, which only involved the verification of positive samples detected by the preliminary deep learning classifier. Although the accuracy of the preliminary classifier was not as high as that reported here for the final classifiers, it nevertheless allowed to significantly speed-up the labeling process by automatically selecting the portions or recordings that most likely contained dolphin whistles.

The human expert was instructed to identify dolphin's whistles as curving lines in the time-frequency domain and to ignore contour lines produced by shipping radiated noise. When discrimination was challenging, the expert directly listened to the recorded audio track to identify whistle-like sounds. The tagging resulted in a binary classification (whistle \textit{vs.} noise) and a contour line marking the time-frequency characteristic of the identified whistle. The latter was used to explore the quality of the manual tagging by checking that the bandwidth of the identified whistle met expected thresholds for a dolphin's whistle, namely between 3~kHz to 20~kHz. A second quality assessment was made by measuring the variance of the acoustic intensity of the identified whistle along the time-frequency contour, where we expect the acoustic intensity of a valid whistle to be stable.

\subsection{Baseline detection method}
As a benchmark detection method we used PamGuard \citep{gillespie2013automatic}, which is a popular software specifically developed to automatically identify vocalizations of marine mammals. The working parameters of PamGuard were set as follows:
\begin{itemize}
\item the ``Sound Acquisition" module from the ``Sound Processing" section was added to handle a data acquisition device and transmit its data to other modules;
\item the ``FFT (spectrogram) Engine" module from the ``Sound Processing" section was added to compute spectrograms;
\item the ``Whistle and Moan Detector" module from the ``Detectors" section was added to capture dolphin whistles;
\item the ``Binary Storage" module from the ``Utilities" section was added to store information from various modules.
\item  the ``User Display" module from the ``Displays" section was added to create a new spectrogram display.
\end{itemize}

Input spectrograms were created using the FFT analysis described above, using the same parameters: FFT window length was set to 2048 points, and the hop size was set to the length multiplied by 0.8 using the Blackman window from the ``FFT (spectrogram) Engine" module in the software settings. The frequency range was set between 3 and 20 kHz, and ``FFT (spectrogram) Engine Noise free FTT data" was selected as source of FFT data from the ``Whistle and Moan Detector" module in settings. While creating a new spectrogram display, the number of panels was set to 2 to visualize both channels. The PamGuard output was considered as a true positive detection if the signal window identified by the software overlapped with at least 5\% of the ground truth signal interval. Although this might seem a permissive criterion, it allowed to consider many PamGuard detections that otherwise would have been discarded.

\subsection{Deep learning detection methods}
We explored two different deep neural network architectures: a vanilla CNN and a pre-trained CNN based on the VGG16 architecture used in object recognition \citep{simonyan2014very}. Note that the spectrogram images were resized to 224x224 and converted into 3D tensors in order to match the number of input channels required by VGG. This was simply achieved by replicating the same image array across the 3 dimensions. Image pixels were normalized by dividing each RGB value by 255.

The vanilla CNN model included two convolutional layers interleaved with max pooling layers (pool size = 2) and dropout layers (dropout factor = 0.2). The convolutional layers used 16 and 32 kernels, respectively, with kernel sizes of (7,7) and (5,5), and a stride value of 2. The last convolutional layer was then flattened and fed to 2 fully connected layers containing 32 and 16 nodes, respectively. All layers used a ReLU activation function; only the output layer used a softmax activation. The model was trained using the Adam optimizer with an initial learning rate of 0.0001.

To implement the transfer learning architecture, the top layers of a pre-trained VGG16 were replaced by 2 new fully connected layers with size 50 and 20, respectively, and the \textit{trainable parameter} was set to ``True''. This allowed the optimizer to jointly train all layers of the VGG model, in order to also adjust low-level features to the new data domain. A ReLU activation function was used in both fully connected layers, while the output layer used a softmax activation. The model was trained using the Adam optimizer with an initial learning rate of 0.00001.

In both cases, binary cross-entropy was used as a loss function and overfitting was monitored by using an early stop criteria (with patience parameter of 15 epochs) applied to a separate validation set. Deep learning models were implemented using Tensorflow \citep{abadi2016tensorflow}. All model hyperparameters were automatically optimized using the Optuna framework (\url{https://optuna.org/}).

\subsection{Evaluation procedure}
To guarantee a robust assessment of our detection method, the dataset was split into separate training and test sets. The training set only contained spectrograms obtained from audio files recorded between July 24th and July 30th, while the test set only contained spectrograms of audio files recorded between July 13th and July 15th. This allowed to test the generalization performance of our models using a completely different set of recordings, thus evaluating the detection accuracy with variable sea conditions.
Overall, the training set contained 108317 spectrograms, of which 49807 were tagged as noise and 58510 as dolphin whistles. The test set contained 6869 spectrograms, of which 4212 were tagged as noise and 2657 were tagged as dolphin whistles. The training set was then randomly shuffled and further split into training and validation sets, using 5-fold cross-validation. Cross-validation was implemented using the ``StratifiedKFold'' function from the scikit-learn library in order to make sure that each validation set contained a balanced amount of data from both classes.

Model performance was assessed by computing mean detection accuracy and by visualizing confusion matrices. True Positive rate and False Positive rate were also computed in order to produce Receiver Operating Characteristic (ROC) curves and measure the corresponding Area Under the Curve (AUC) \citep{davis2006relationship}:
\begin{equation} 
\label{eq:PrecRec}
\begin{array}{l}
Precision=
 \frac{TP}{TP+FP};\\
 \vspace{3pt}
Recall=
 \frac{TP}{TP+FN};\\
 \vspace{3pt}
True \: Positive \: Rate=
 \frac{TP}{TP+FN};\\
 \vspace{3pt}
False \: Positive \: Rate=
 \frac{FP}{FP+TN}
\end{array}
\end{equation}
where $TP$ indicates True Positives, $TN$ True Negatives, $FP$ False Positives and $FN$ False Negatives.

\section{Results}

The vanilla CNN model achieved a remarkable mean detection accuracy of 80.6\%, significantly outperforming the PamGuard baseline, which achieved 66.4\%. Most notably, the performance of the VGG model implementing the transfer learning approach was even more impressive, achieving a mean detection accuracy of 92.3\%.

\begin{figure}[]
    \centering
    \includegraphics[width=0.9\textwidth]{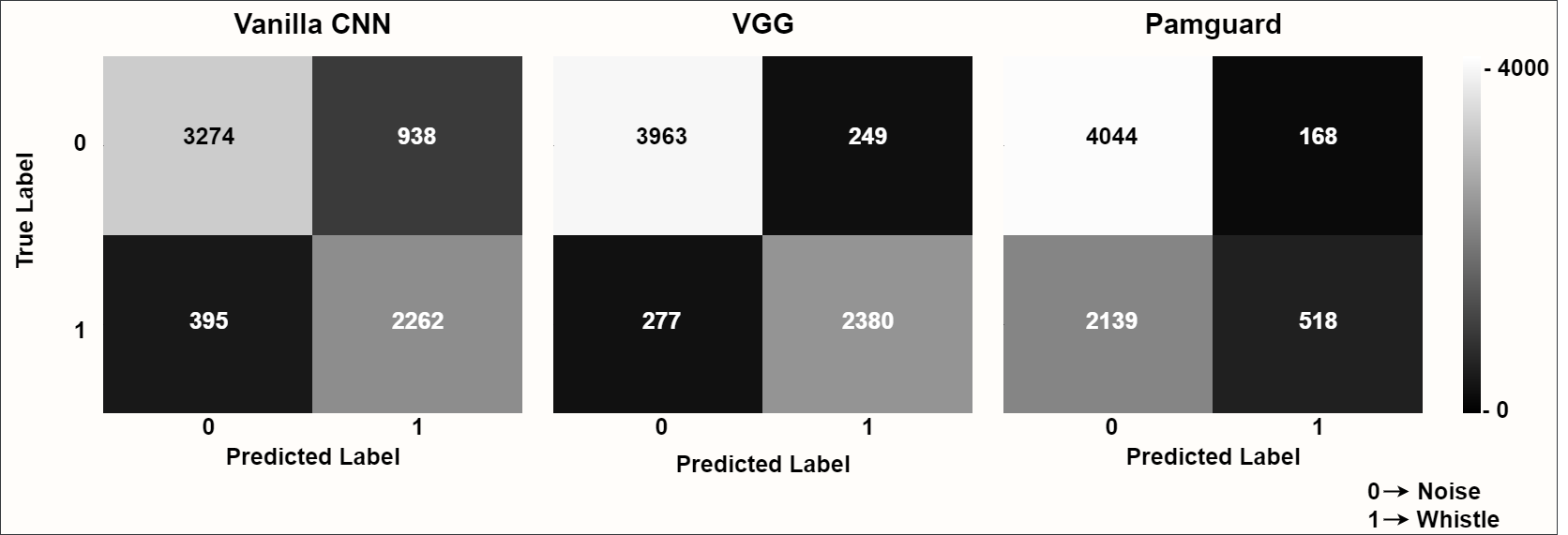}
    \caption{Confusion matrices for the deep learning models (vanilla CNN and VGG with transfer learning) and for the PamGuard baseline.}
    \captionsetup{justification=centering}
    \label{fig:CM}
\end{figure}

The advantage of deep learning models is even more striking when considering the confusion matrices: as shown in Fig.~\ref{fig:CM}, although the amount of True Negatives (Label = 0) was comparable across different methods, the number of True Positives was remarkably higher for deep learning models, especially for the VGG architecture. The low sensitivity of PamGuard was highlighted by the very high number of False Negatives ($n$ = 2139), suggesting that this method is not very effective in identifying dolphin whistles when the level of signal to noise ratio makes detection particularly challenging.

\begin{figure}[]
    \centering
    \includegraphics[width=0.8\textwidth]{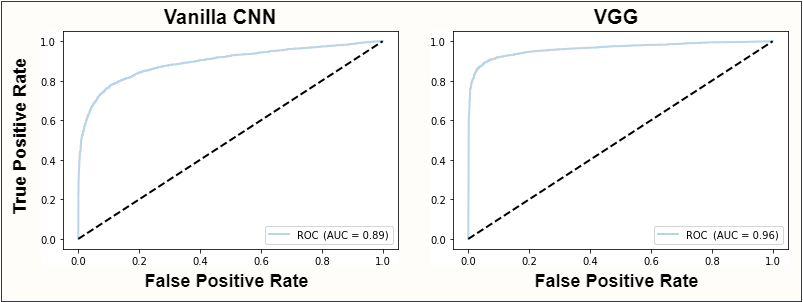}
    \caption{Receiver Operating Characteristic (ROC) curves and corresponding Area Under the Curve (AUC) values for both deep learning models.}
    \captionsetup{justification=centering}
    \label{fig:ROC}
\end{figure}

The ROC curves and AUC scores reported in Fig.~\ref{fig:ROC} allow to further compare the performance of deep learning models. The superior accuracy of the VGG architecture is evident also in this case, approaching the performance of the ideal classifier.

\section{Discussion}
With the large increase in human marine activity, our seas have become populated with boats and ships projecting acoustic emissions of extremely high power that often affect areas of more than 20 $\mathrm{km}^2$. The underwater radiated noise level from large ships can exceed 100 PSI with a clear disturbance impact on the hearing, self-navigation and foraging capabilities of marine mammals and especially coastal dolphins \citep{ketten2008underwater, erbe2019effects}. Monitoring the marine ecosystem and the sea life is thus a crucial task to promote environment preservation.

Nevertheless, traditional monitoring technologies rely on sub-optimal detection methods, which limit the possibility of conducting long-term and large-scale surveys. Automatic detection methods can greatly improve our surveying capability, however algorithmic solutions do not achieve satisfactory performance in the presence of high levels of background noise. In this paper we demonstrated that modern deep learning approaches can detect dolphin whistles with an impressive accuracy, and are thus well-suited to become the new standard for the automatic processing of underwater acoustic signals. Although further research is needed to validate these methods in different marine environments and with different animal species, we believe that deep learning will finally enable the creation and deployment of cost-effective monitoring platforms.

\section*{Conflict of Interest Statement}
The authors declare that the research was conducted in the absence of any commercial or financial relationships that could be construed as a potential conflict of interest.

\section*{Author Contributions}
RD and AT contributed to conception and design of the study. RD performed the sea experiments and provided the recordings database. BKN and AT designed the deep learning models. BKN implemented the models and performed the analyses. GD performed data tagging. All authors contributed to manuscript writing, revision, read, and approved the submitted version.

\section*{Funding}
The research was funded in part by a grant from the University of Haifa’s Data Science Research Center.

\section*{Data Availability Statement}
The datasets created and analyzed for this study can be found here: https://csms-acoustic.haifa.ac.il/index.php/s/2UmUoK80Izt0Roe

\bibliography{bibliography.bib}

\end{document}